\documentclass{article}
\usepackage{graphicx}
\newtheorem{fig}{Figure}

\newtheorem{thm}{Theorem}

\newtheorem{tab}{Table}

\def\ligne#1{\hbox to \hsize{#1}}

\def\leurre{\noindent\leftskip0pt\small\baselineskip 10pt}
\def\grostrait{\ligne{\vrule height 1pt depth 1pt width \hsize}}
\def\demitrait{\ligne{\vrule height 0.5pt depth 0.5pt width \hsize}}

\title{
Coordinates for a new triangular tiling of the hyperbolic plane
}
\author{Maurice Margenstern}
\begin{document}
\maketitle

\begin{center}
Laboratoire d'Informatique Th\'eorique et Appliqu\'ee, EA 3097,\\
        Universit\'e Paul Verlaine $-$ Metz, UFR-MIM, \\
and CNRS, LORIA,\\
        \^Ile du Saulcy, 57045 Metz Cedex, France\\
{\it e-mail}: {\tt margens@univ-metz.fr}
\end{center}

\def\titlerunning{Coordinates for a new triangular hyperbolic tiling}
\def\authorrunning{Margenstern}

\begin{abstract}
In this paper we define an infinite family of triangular tilings of the hyperbolic
plane defined by two parameters ranging in the natural numbers and we give a uniform 
way to define coordinates for locating the triangles of the tiling. 
\end{abstract}

\section{Introduction}
\label{intro}
   In~\cite{mmbook2}, the author introduced a system of coordinates for the points of the
hyperbolic plane. The starting point of the system is a tessellation of the hyperbolic
plane defined by a regular convex polygon~$P$ with $p$~sides and with interior angle
$\displaystyle{{2\pi}\over q}$. This latter expression means that it is possible 
to exactly cover a neighbourhood of a point~$V$ by placing $q$~copies of~$P$ with $V$~as 
a common vertex, each copy being an image of the initial one by a suitable 
rotation around~$V$ involving a multiple of the above angle. In Section~\ref{tilings} we 
explain how to define coordinates for the tiling defined by~$P$. In Section~\ref{trigrids},
remembering the construction of~\cite{mmbook2}, we shall define a family of triangular 
tilings together with a coordinate system for each member of the family.
In Subsection~\ref{coordtrigrids}, we give an algorithm to compute the coordinates of
the neighbours of a tile from the coordinate of this tile.

\section{The tilings $\{p,q\}$}
\label{tilings}

   In order to be as self-contained as possible, we start by reminding in what consists
hyperbolic geometry, also introducing a model in which we can see something, this
is the goal os Sub-section~\ref{introgeom}. However,
it is the place here to warn the reader that despite the model we can see a very 
little only. In fact,we are in the situation of the pilot of a plane with no visibility, 
when the flight has to be processed with instruments only. 

Then, in Sub-section~\ref{grids}, we remind the tessellations of the hyperbolic plane 
constructed from a regular convex polygon and, in Sub-section~\ref{coord} we remember
the definition of the coordinates for the tiles in these tessellations.

\subsection{Hyperbolic geometry}
\label{introgeom}

Hyperbolic geometry appeared in the first half of the
19$^{\rm th}$ century, proving the independence of the parallel
axiom of Euclidean geometry. Models were devised in the second
half of the 19$^{\rm th}$ century and we shall use here one of the
most popular ones, Poincar\'e's disc. This model is represented by
Figure~\ref{poincare_disc}.

\vskip 10pt
\vtop{
\ligne{\hfill
\scalebox{1.00}{\includegraphics{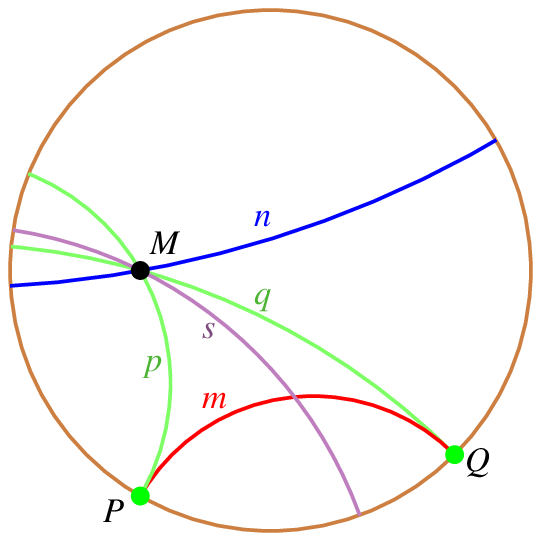}}
\hfill}
\begin{fig}\label{poincare_disc}
Poincar\'e's disc model.
\end{fig}
}

Inside the open disc represented in the figure we have the points of the 
hyperbolic plane. Note that by definition, the points on the border of the disc 
do not belong to the hyperbolic plane. However, these points play an important role
in this geometry and are called {\bf points at infinity}. Lines are trace of 
diameters or circles orthogonal to the border of the disc, e.g. the line~$m$. 
In this model, two lines which meet in the open disc are called {\bf secant}
and two lines which meet at infinity, {\it i.e.} at a point at infinity
are called {\bf parallel}. In the figure, we can see a line~$s$ through 
the point~$A$ which cuts~$m$. Now, we can see that two lines pass through~$A$
which are parallel to~$m$: $p$ and~$q$. They touch~$m$ in the model at~$P$ and~$Q$
respectively which are points at infinity. At last, and not the least: 
the line~$n$ also passes through~$A$ without cutting~$m$,
neither inside the disc nor outside it. This line is called {\bf non-secant}
with~$m$.

\subsection{The tilings $\{p,q\}$}
\label{grids}

From a famous theorem established by Poincar\'e in the late 19$^{\rm th}$ century,
it is known that there are infinitely many tilings in the hyperbolic plane, 
each one generated by the reflection of a regular convex polygon~$P$ in its sides and, 
recursively, by the reflection of the images in their sides, provided that the 
number~$p$ of sides of~$P$ and the number~$q$ of copies of~$P$ which can be put 
around a point~$A$ and exactly covering a neighbourhood of~$A$ without overlapping 
satisfy the relation:
\hbox{$\displaystyle{1\over p}+\displaystyle{1\over q}<\displaystyle{1\over2}$}.
The numbers $p$ and~$q$ characterize the tiling which is denoted $\{p,q\}$ and
the condition says that the considered polygons live in the hyperbolic plane. 
We call~$P$ the {\bf basic polygon} of the tessellation. Note
that the three tilings of the Euclidean plane which can be defined up to similarities
can be characterized by the relation obtained by replacing~$<$ with~$=$ in the above
expression. In this way, the basic polynomial of these tessellations are
the square for $\{4,4\}$, the equilateral triangle for $\{3,6\}$and
the regular hexagon for $\{6,3\}$.

   In the paper, the figures which illustrate the explanations of the text are
based on two tilings which are the simplest ones which can be defined in this way in the 
hyperbolic plane: $\{5,4\}$ and~$\{7,3\}$. We call these tilings the {\bf pentagrid}
and the {\bf heptagrid} respectively. They are illustrated by Figures~\ref{tiling_54}
and~\ref{tiling_73}. On the right-hand side of the figure, we can see the tree which 
is in bijection with an angular sector, a basic structure of the heptagrid, also see
Figure~\ref{eclate_73}. 
\vskip 10pt
\vtop{
\ligne{\hfill
\scalebox{0.28}{\includegraphics{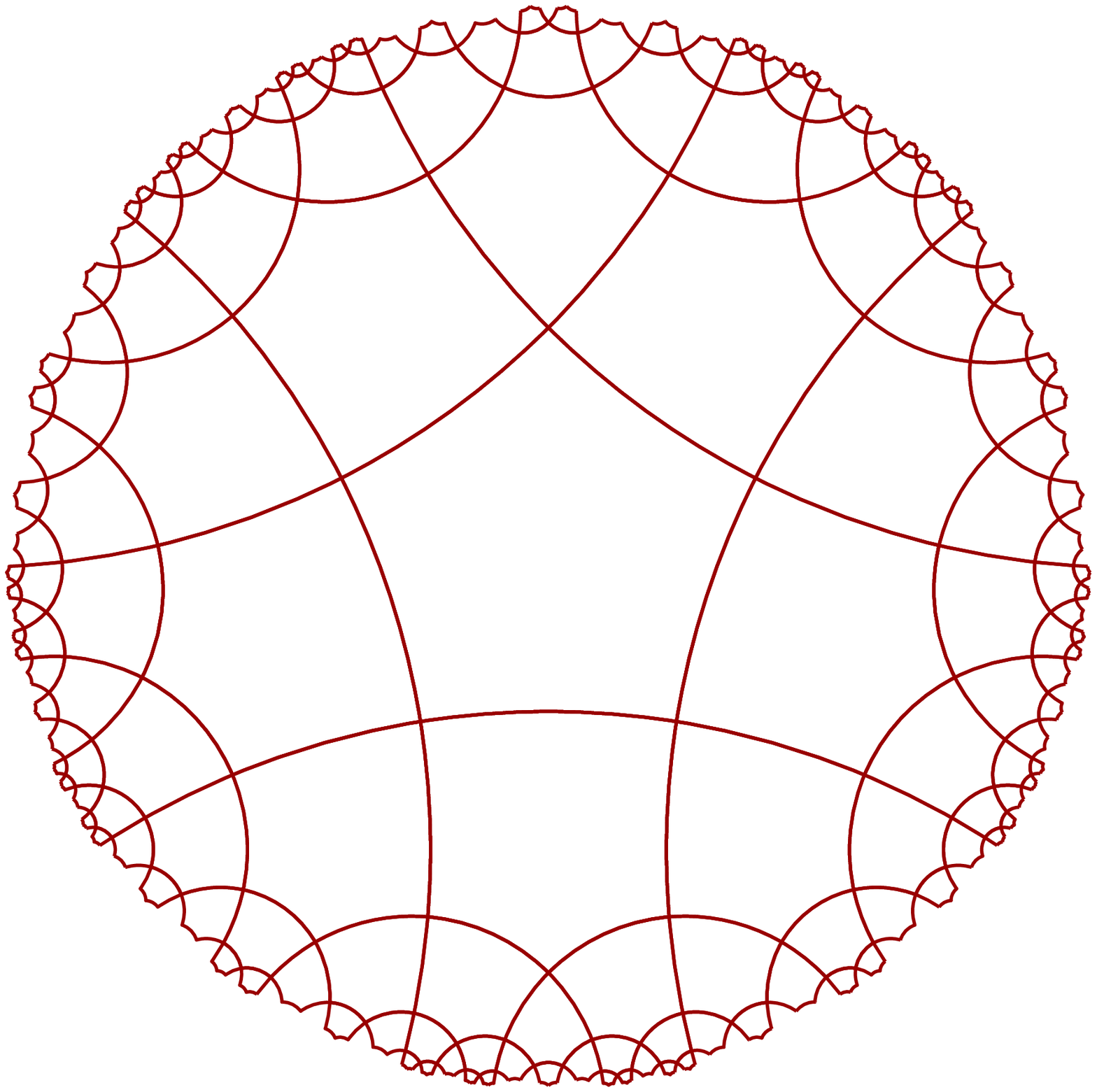}}
\scalebox{0.975}{\includegraphics{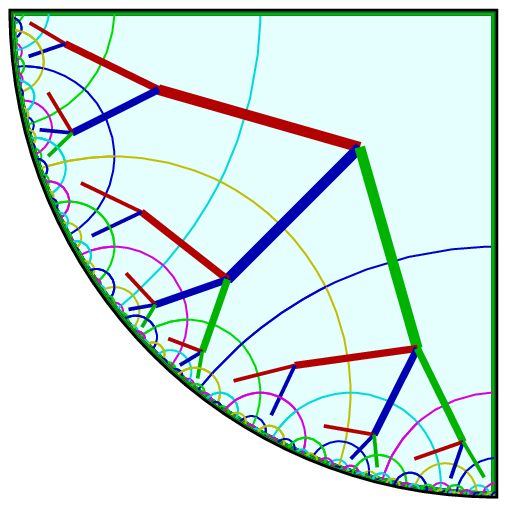}}
\hfill}
\begin{fig}\label{tiling_54}
The pentagrid. On the right-hand side: the key structure to explore the tiling.
\end{fig}
}

\vskip 10pt
\vtop{
\ligne{\hfill
\scalebox{0.40}{\includegraphics{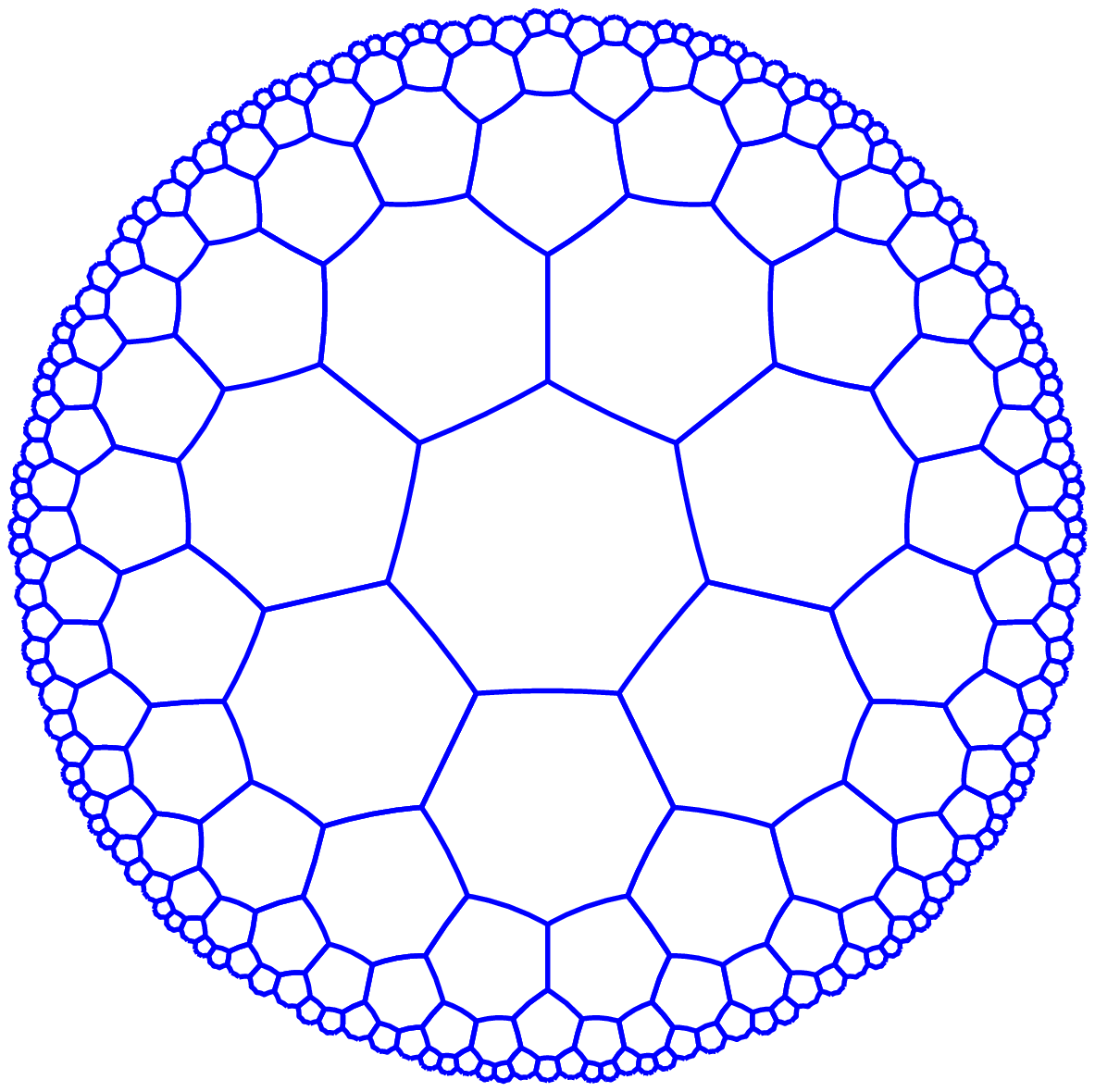}}
\scalebox{1.00}{\includegraphics{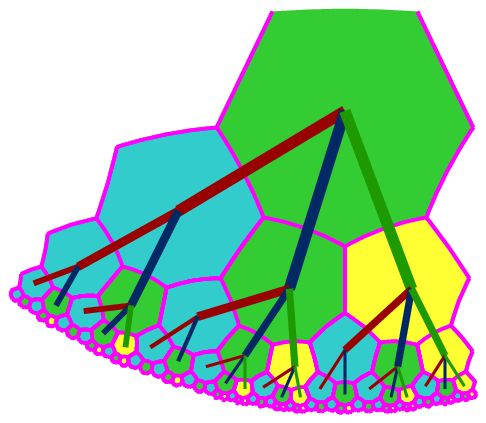}}
\hfill}
\begin{fig}\label{tiling_73}
The heptagrid. On the right-hand side: the key structure to explore the tiling.
\end{fig}
}

   This property is not specific to the heptagrid, it is a general property of the
tilings $\{p,q\}$. We refer the reader to~\cite{mmbook1,mmbook2,mmarXivnewpq} for a 
detailed analysis and detailed explanations of these tools. However, here we give a
summary of these properties which allow the reader to better understand the further
constructions.

   The angular sector which we put in bijection with a tree is defined in a way
which depends on the parity of~$q$.

   When $q$ is even, we fix a copy~$P$ of the basic polygon of the tessellation.
We fix a vertex~$V$ of~$P$ and we consider~$\ell$ and~$r$ the
two rays issued from~$V$ which support the sides of~$P$ ending at~$V$. If we turn 
around~$V$ in the counter-clockwise orientation and remaining in~$P$, we can first 
see~$\ell$ and then~$r$. The angle defined by~$V$, $\ell$ and~$r$ taken in this order
in the above mentioned orientation, constitutes an angular sector and we
say that $P$~is its {\bf leading tile} and that $V$~is its {\bf summit}. As proved 
in~\cite{mmbook1}, the restriction of
the tiling to this angular sector is in bijection with a tree, and
this tree looks like the one which illustrated by Figure~\ref{tiling_54} when $q=4$. 

   When $q$ is odd, we again fix a copy~$P$ of the basic polygon to be the central cell. 
However, this time we cannot take the sides which abut to a vertex in order to define
a sector. The reason is that the line which supports this side cuts another tile along
one of its reflection axes and this leads to further complications. A simpler way
is the following one, explained in~\cite{mmarXivnewpq}. It is illustrated by
Figure~\ref{vertex}. 

First, we define the type of lines which we shall use to delimit the angular sector.
Consider a copy of~$P$ and 
fix a vertex~$V$ of~$P$. We know that exactly $q$~sides of copies of~$P$ have~$V$ 
as an end. There is a single one~$e$ which is supported by the bisector of the angle 
defined by the two edges of~$P$ which meet at~$V$. Consider the mid-point~$M$ of~$e$.
Consider the edge of a copy of~$P$ 
which makes the angle
$\vartheta=\displaystyle{{2\pi.h}\over q}$ with~$e$, the angle being measured while 
turning counter-clockwise around~$V$. The over end of this edge is~$W$. Next, repeat 
the construction from~$W$ with the just considered edge and considering the edge 
defined by the angle~$\vartheta$ with~$VN$ measured while turning around~$W$ clockwise 
this time. This defines a new edge whose
other end is~$X$. Now, the mid-points $M$, $N$ and~$O$ of the just considered edges
are on a same ray~$\delta$ issued from~$M$. Next, we can endlessly repeat these 
two steps of construction starting from~$X$. The line which support this ray is called a 
\hbox{\bf $h$-mid-point line} and the ray itself is a {\bf $h$-mid-point ray}. Note 
that another $h$-mid-point ray is issued from~$M$: it is obtained by taking the reflection 
of~$\delta$ in the line which supports~$e$.

Second, we can now define the angular sector as follows. We fix a copy of~$P$ which
is the leading tile of the sector. Let $b$ and~$c$ be the edges of~$P$ which meet 
at~$V$. We may assume that while counter-clockwise turning around~$V$ $c$~is met 
before~$b$. Let $a$~be the edge which has~$V$ as an end and which is supported by the
bisector of the angle defined by~$b$ and~$c$. Let~$B$ and~$C$ be the mid-points of~$b$
and~$c$ respectively. Let~$r$ be the $h$-mid-point ray issued from~$M$ which
passes through~$B$. Consider the rotation~$\rho$ around~$V$ which transforms~$b$ 
into~$c$: it leaves globally invariant the set of edges meeting at~$V$. Accordingly,  
$\rho$~transforms~$M$ into~$M$', the mid-point of another edge meeting at~$V$ which is
on the bisector of the angle defined by~$c=\rho(b)$ and~$\rho(c)$. It is easy to
see that the $h$-mid-point ray~$\ell$ issued from~$M'$ which passes through~$C$ is 
exactly $\rho(r)$, see Figure~\ref{sect_S0}.
The angular sector which we define has its summit at~$V$, its leading tile is~$P$
and it is delimited by the $h$-mid-point rays~$\ell$ and~$r$.
Now, the set of tiles we consider as attached
to this angular sector is the set of tiles whose all vertices but possibly at most one
are included in the sector, its two rays being included. We again say that this set
of tiles is the restriction of the tiling to the angular sector.

\vtop{
\ligne{\hfill
\scalebox{0.8}{\includegraphics{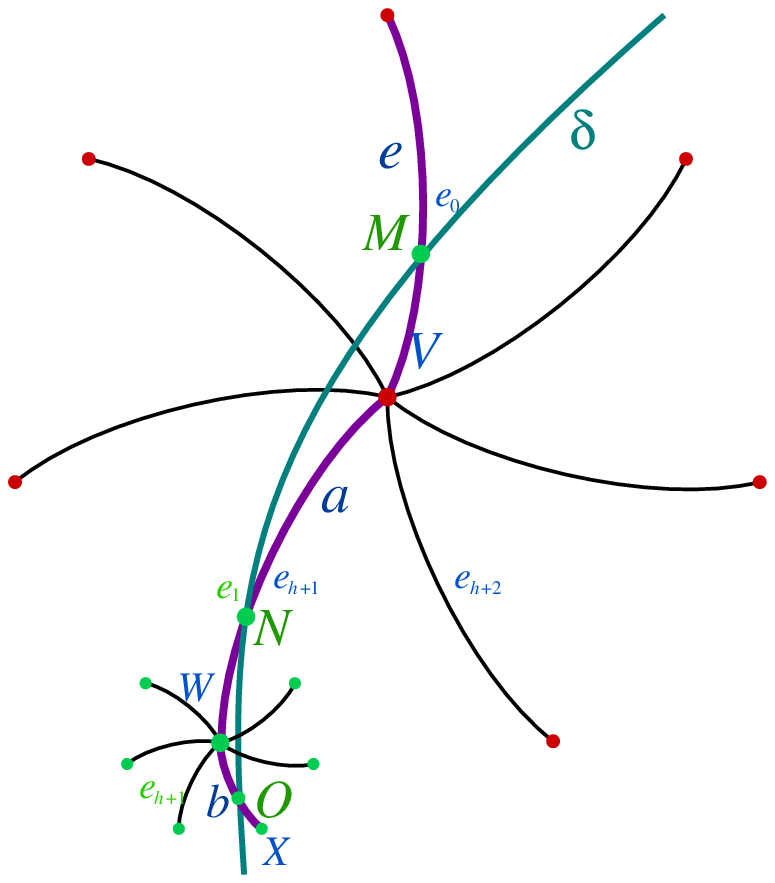}}
\hfill}
\vspace{-30pt}
\begin{fig}
\label{vertex}\small
Illustration of the construction of the $h$-mid-point line.
\end{fig}
}

\vtop{
\ligne{\hfill
\scalebox{0.8}{\includegraphics{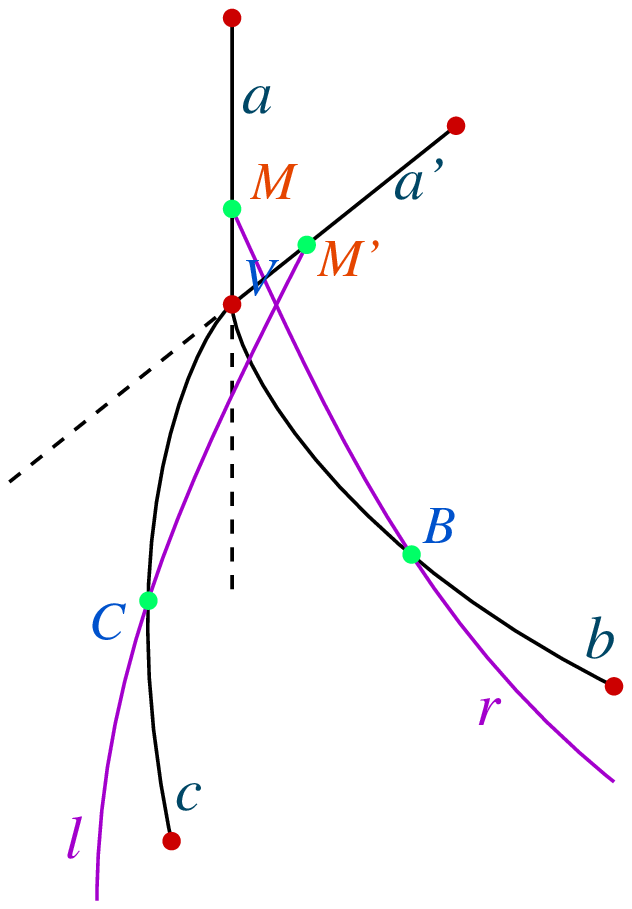}}
\hfill}
\vspace{-10pt}
\begin{fig}
\label{sect_S0}\small
The $h$-mid-point rays used for the definition of the angular sector when $q$~is odd. 
The rotation
around~$V$ which transforms~$r$ into~$\ell$ also transforms~$M$ into~$M'$ which
is the mid-point of an edge abutting~$V$ of a copy of~$P$ around~$V$.
\end{fig}
}

   Now, in both cases, the whole tiling splits into the central cell and 
\hbox{$p.(h$$-$$1)$} copies of the angular sector dispatched around the central tile.

   It is important to note that the tree which is obtained in this case
is a bit different from the tree which we obtain in the case when $q$~is even. 

   Before turning to the coordinates, we have to mention two important particular
cases: the case when $q=4$ and the case when $q=3$. In both of them and in these 
cases only, the tilings $\{p,4\}$ and $\{p$+$2,3\}$ share very particular properties.

   The first property is that for each $p\geq 5$, the tilings $\{p,4\}$ and
$\{p$+$2,3\}$ have their angular sector spanned by the same tree.

   The second property shared by all tilings $\{p,4\}$ and $\{p$+$2,3\}$ is the 
following. Fix a copy of~$P$, 
the basic polygon of the tessellation. We call it the {\bf central tile}. Say that
$P$~belongs to generation~0. Now, the tiles obtained by reflection of a polygon
of the generation~$n$ in its sides and which do not belong to a generation~$m$
with~$m\leq n$ are said to belong to the generation~$n$+1. Now from each tile of
generation~1, we can define a copy of an angular sector, taking this tile as
the leading tile of the sector, in such a way that the sectors and the central tile
cover the whole plane without overlapping. 

This can be obtained as follows in the
case when $q=4$. The
summit of each sector is a vertex of the central tile~$C$. Let $P$~be the leading tile
of the angular sector. We place the sector in such a way that the side shared by~$P$
and~$C$ is supported by the ray~$\ell$ of the sector. We can see that the continuation
of~$r$ also supports a side of~$C$, but this side is not contained in the sector: it is
contained in another one. 

In the case when $q=3$, the $h$-mid-point lines coincide with 
the {\bf mid-point lines} defined as lines joining mid-points of consecutive edges in
a polygon. Then, the angular sector is defined by taking a neighbour~$P$ of the central
tile as leading tile of the sector. We chose~$V$ a vertex belonging to~$P$ and to the
central tile in such a way that the ray~$\ell$ passes through the mid-point of the 
common edge of~$P$ and the central tile. 

In both cases, it is not difficult to see that we have $q$~angular sectors
around the central tile and that the sectors and the central cell cover the plane without
overlapping, see Figure~\ref{dispatch} for the pentagrid and the heptagrid.
The levels of the tree which is in bijection with the
restriction of the tiling to the sector define levels in the sector, thanks to the
bijection. The leading tile is the root of the tree, whose level is~0, and it is not 
difficult to see that the level~$n$+1{} in the sector consists of the restriction, to
the sector, of the tiles of the generation~$n$. 

\vskip-25pt
\vtop{
\ligne{\hfill
\scalebox{0.45125}{\includegraphics{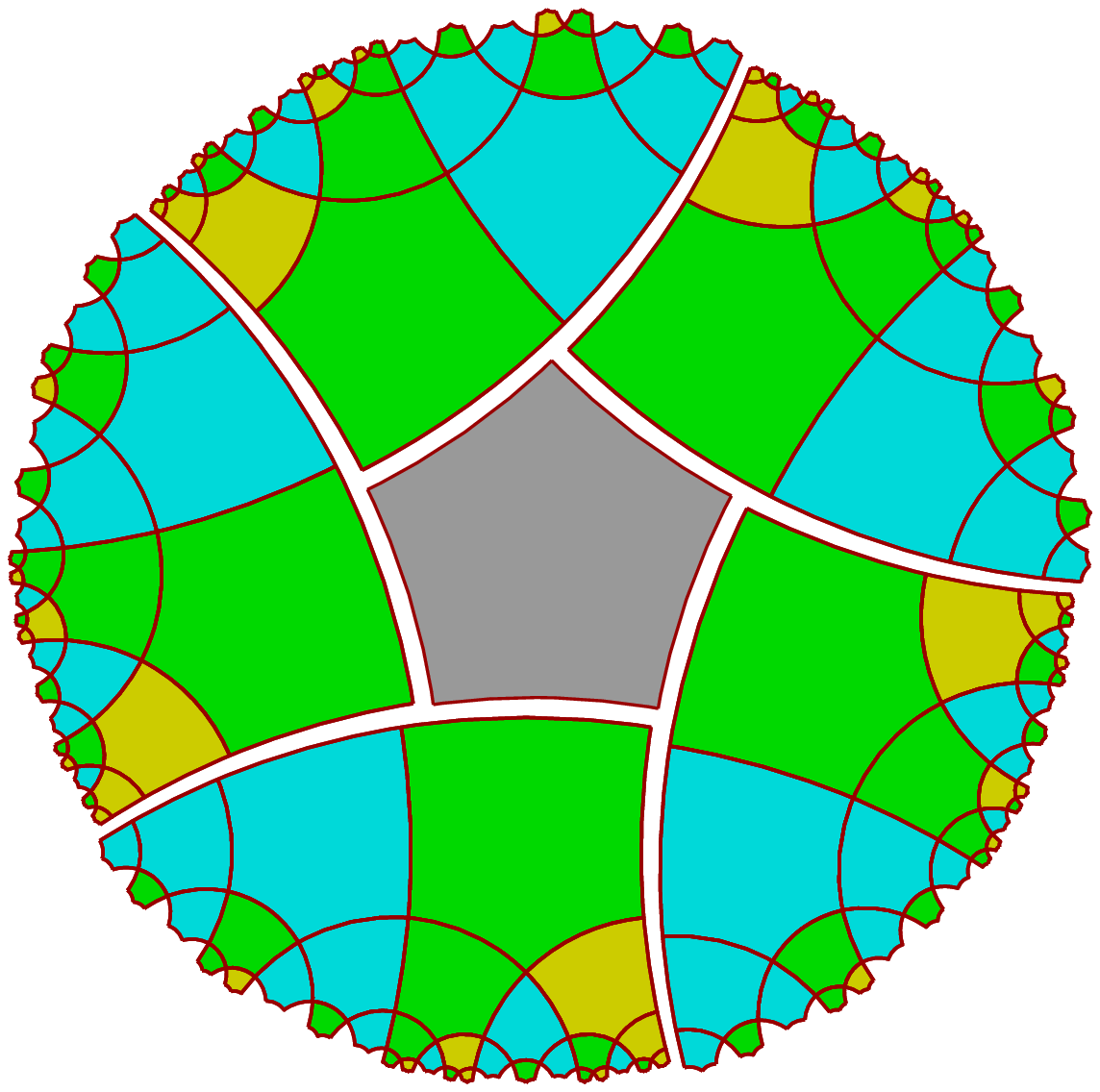}}
\hskip-20pt
\raise 35pt\hbox{\scalebox{0.965675}{\includegraphics{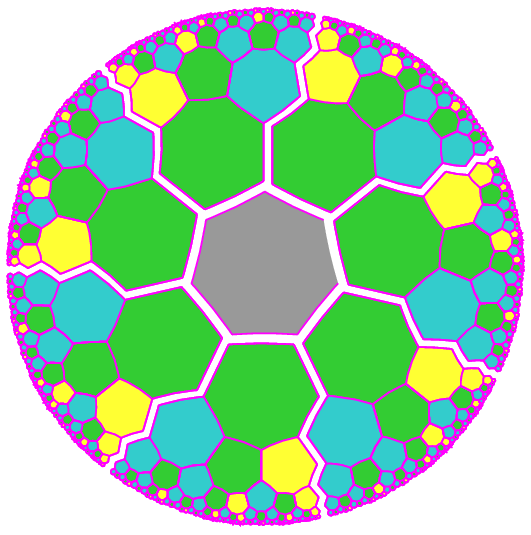}}}
\hfill}
\vspace{-35pt}
\begin{fig}
\label{dispatch}\small
The display of $q$~sectors around the central tile.
Left-hand side: in the pentagrid. Right-hand side: in the heptagrid.
\end{fig}
}

   This correspondence between the generation by reflection in the sides and the
levels of the tree is no more true when $q > 4$. This is due to the fact that, starting 
from $q=6$ in the case when $q$~is even and from $q=5$ in the case when $q$~is odd, one 
level of the tree contains tiles belonging to several generations, but not
all of them, so that the correspondence between the levels and the generations is much 
more complex.

\subsection{Coordinates for the tilings $\{p,q\}$}
\label{coord}

The existence of a tree which is in bijection with the restriction of the
tiling to an appropriate angular sector is a key property. It is the basis of a 
very efficient navigation tool to locate tiles in these tilings. This tool is even 
more efficient in the cases of the pentagrid and of the heptagrid.

   In the case when $q$~is even, the coordinates of the tiles are obtained by the 
greatest positive root of the following polynomial: 

\ligne{\hfill
$P(X) = X^2-((p$$-$$3).(h$$-$$1)$$+$$1).X - h+3$,
\hfill} 

\noindent
where
$h$~is defined by $q=2h$. 
There are two kind of nodes: ordinary ones with 
\hbox{$(p$$-$$3).(h$$-$1$)$+1} nodes
and special ones with \hbox{$(p$$-$$2).(h$$-$1$)$$-$1}~nodes. 
In the case when $q$~is odd the coordinates are obtained from the greatest real
root of another polynomial:

\ligne{\hfill
$P(X) = X^3 - ((p$$-$$3)(h$$-$$1)$+$1)X^2 - ((p$$-$$2)(h$$-$$1)$$-$$2)X - h$+3,
\hfill}

\noindent
where $h$~is defined by $q=2h$+1.

This time, the number of sons for the ordinary nodes is 
\hbox{$(p$$-$$3).(h$$-$1$)$+2} nodes
and among them, we have two nodes with special types, say~0 and~1, one node in each type. 
For the special nodes of type~0 they have \hbox{$(p$$-$$3).(h$$-$1$)$+1} sons where 
one of them is of type~1 and all others are ordinary nodes.
For the special nodes of type~1 they have
\hbox{$(p$$-$$2).(h$$-$1$)$$-$1}~nodes with a single node of special type among them, 
namely the type~1.

    In both cases, the greatest real root~$\beta$ of $P(X)$ is positive.
In fact it satisfies the following relation: 
\hbox{$\beta > (p$$-$$3)(h$$-$$1)$+1} when $q$~is even and
\hbox{$\beta > p$$-$2} when $q$~is odd, assuming in both cases that
$p\geq 5$ and $h\geq 4$. Now, for the remaining three cases, easy and direct computations 
show that \hbox{$\beta > 1$}. 
From~$P(X)$, we define an induction equation: 

\ligne{\hfill
$u_{n+2} = ((p$$-$$3).(h$$-$$1)$$+$$1).u_{n+1} + (h$$-$$3).u_n$,
\hfill} 

\noindent
for even~$q$ with $u_0=1$ and \hbox{$u_1=(p$$-$$3).(h$$-$$1)$$+$$1$}. When $q$~is odd,
we define:

\ligne{\hfill
$u_{n+3} = ((p$$-$$3)(h$$-$$1)$+$1).u_{n+2} + ((p$$-$$2)(h$$-$$1)$$-$$2).u_{n+1} + 
(h$$-$$3).u_n$,
\hfill}

\noindent
with $u_0=1$, $u_1=(p$$-$$3)(h$$-$$1)$+$2$ and 
$u_2=((p$$-$$3)(h$$-$$1))^2+(4p$$-$$11)(h$$-$$1)$. Now we can notice that
if we take $u_{-1}=0$, we obtain that:\vskip 0pt
\ligne{$u_2= ((p$$-$$3)(h$$-$$1)$+$1).u_1
+((p$$-$$2)(h$$-$$1)$$-$$2).u_0 + (h$$-$$3).u_{-1}$.\hfill}

We have that $u_n$ is the number of nodes of the level~$n$ in the tree~$\cal T$ in 
bijection with the restriction of the tiling to the angular sector.

Introducing $b=\lfloor\beta\rfloor$,
it is known that any positive natural number~$n$ can be represented as a sum of
the form $n=\displaystyle{\sum\limits_{i=0}^ka_iu_i}$, where $\{u_n\}_{n\in I\!\!\!N}$
is defined by the equation associated to~$P(X)$. In general, the representation is 
not unique. However, it can be made unique by requiring the representation which has the
greatest number of digits, or in terms of the above formula, where $k$~is maximal.
We call this representation the {\bf greedy} representation of~$n$ in the 
basis~$\beta$. Numbering the nodes of~$\cal T$ from the root and then level by level 
and, on each level from the left to the right, we call {\bf coordinate} of a 
tile~$T$ of an angular sector the greedy representation of the number attached 
to~$T$ under the bijection between~$\cal T$ and the sector. Now, the central tile
has 0~as coordinate and any other tile is in a single sector~$\sigma$ with
$\sigma\in[1..max]$, where $max=p(h$$-$$1)$. Accordingly, we call {\bf coordinate}
of a tile~$T$, either 0, for the central tile, or $(\sigma,\nu)$ where $\sigma$~is
the number of the sector containing~$T$ and $\nu$~is its coordinate in~$\cal T$.

   Now, the language of the coordinates is regular in all cases but one. The reason
is the following. It is not difficult to see that the other
roots of~$P(X)$ have a modulus less than~1 so that we always have a Pisot polynomial,
except in the case when $q=5$ and $p=4$. Now, it is known that
the language of the greedy representation of the positive natural numbers in the
basis~$\beta$ is regular if and only if $\beta$~is the greatest real root of a Pisot
polynomial. However, in the exceptional case, we can introduce coordinates using the 
dual graph of $\{4,5\}$ which is the pentagrid: as the pentagrid has a coordinate 
system with much more convenient properties we can use this system for 
$\{4,5\}$, see~\cite{mmarXivnewpq} for more details.
 
    We refer the reader to~\cite{mmbook1} for a detailed study of the properties
of the coordinates of the tiles which allows us there to obtain the computation of 
the neighbours of a tile. Remember that by neighbour of a tile~$T$, we mean a tile 
which shares a side with~$T$.

We also attach numbers to the sides of a tile~$T$. Assuming that side~1 is fixed,
the other sides are obtained in increasing number while counter-clockwise turning
around~$T$. If~$T$ is the central tile, side~1 is fixed once and for all. If~$T$
is another tile, side~1 is the side shared with its father, assuming that the father
of the leading tile of a sector is the central tile. This numbering of the sides
allows to number the neighbours of a tile~$T$: by definition, the neighbour~$i$ of~$T$
is the tile which shares with~$T$ the side numbered~$i$ in~$T$.

For the following sections, we assume that we have the function $v(\sigma,\nu,\tau)$ 
at our disposal, which computes the neighbour~$\tau$ of the tile~$T$ whose coordinate is
$(\sigma,\nu)$. Most often,
the sector of~$v(\sigma,\nu,\tau)$ is again the sector~$\sigma$. Now, if $\nu$~is
the coordinate of a node on an extremal branch of the tree, the sector 
of~$v(\sigma,\nu,\tau)$ is different: if may be $\sigma\oplus$1 or $\sigma\ominus$1,
where \hbox{$\sigma\oplus$$1=\sigma$+1} when $\sigma\in[1..max$$-$$1]$ and
\hbox{$max\oplus$$1=1$}; similarly, 
\hbox{$\sigma\ominus$$1=\sigma$$-$1}
when $\sigma\in[2..max]$ and \hbox{1$\ominus$$1=max$}. Then, when the sector 
of $v(\sigma,\nu,\tau)$ is not~$\sigma$ it is either $\sigma\oplus$1 or
$\sigma\ominus$1.

   However, we give in Tables~\ref{coord2} and \ref{coord3}
the possibility to compute $v(\sigma,\nu,\tau)$ in the case of the heptagrid in terms 
of two auxiliary functions: $f(\nu)$, which gives the number of the father of a node 
belonging to the tree, and $\sigma(\nu)$ which is the number of neighbour~4{} in all
the cases. As in most cases, the sector of $v(\sigma,\nu,\tau)$ is also~$\sigma$,
we shall consider only the number of the tree for~$v(\nu,\tau)$, the neighbour~$\tau$
of the node~$\nu$. In Table~\ref{coord3}, we mention the computation of the new sector
when the neighbour is outside the sector of the tile.  

\ifnum 1=0 {
\def\lignetab #1 #2 #3 #4 #5 {%
\ligne{\hfill\hbox to 300pt{
\hbox to 80pt{\hfill#1\hfill}
\hbox to 50pt{\hfill#2\hfill}
\hbox to 50pt{\hfill#3\hfill}
\hbox to 50pt{\hfill#4\hfill}
\hbox to 50pt{\hfill#5\hfill}
}
\hfill}
}
\vtop{
\begin{tab}\label{coord1}
\leurre 
Coordinates of the neighbours of a $2$-triangle~$T$ in terms of the coordinates 
of~$T$.
\end{tab}
\vspace{-8pt}
\grostrait
\lignetab {neighbour} {sector} {number} {slice} {place}
\vspace{-3pt}
\demitrait
\lignetab {0} {$\sigma$} {$\nu$} {$\tau$} 0
\lignetab {1} {$\sigma$} {$\nu$} {$\tau$} 3
\lignetab {2} {$s(\sigma,\nu,\tau)$} {$v(\tau,\nu)$} {$t(\nu,\tau)$} 1
\lignetab {3} {$\sigma$} {$\nu$} {$\tau\ominus1$} 1
\vspace{-3pt}
\demitrait
\lignetab {0} {$\sigma$} {$\nu$} {$\tau$} 1
\lignetab {1} {$s(\sigma,\nu\tau)$} {$v(\tau,\nu)$} {$t(\nu,\tau)$} 0
\lignetab {2} {$\sigma$} {$\nu$} {$\tau$} 3
\lignetab {3} {$\sigma$} {$\nu$} {$\tau\oplus1$} 0
\vspace{-3pt}
\demitrait
\lignetab {0} {$\sigma$} {$\nu$} {$\tau$} 2
\lignetab {1} {$\sigma$} {$\nu$} {$\tau\ominus1$} 2
\lignetab {2} {$\sigma$} {$\nu$} {$\tau\oplus1$} 2
\lignetab {3} {$\sigma$} {$\nu$} {$\tau$} 3
\vspace{-3pt}
\demitrait
\lignetab {0} {$\sigma$} {$\nu$} {$\tau$} 3
\lignetab {1} {$\sigma$} {$\nu$} {$\tau$} 0
\lignetab {2} {$\sigma$} {$\nu$} {$\tau$} 1
\lignetab {3} {$\sigma$} {$\nu$} {$\tau$} 2
\vspace{-3pt}
\demitrait
\vskip 7pt
}
}\fi
\vskip 5pt
The computation of these expressions can be found in~\cite{mmbook1,mmbook2}
but we repeat them for the self-containedness of the paper.
The computation of~$v(\tau,\nu)$ involves the already mentioned
auxiliary functions, $f(\nu)$ and~$\sigma(\nu)$. 
As mentioned 
from~\cite{mmkmTCS,mmJUCSii}, the tree which we consider has two kinds of nodes:
black and white ones with two and three sons respectively. The sons can be
deduced from the node by the following rules \hbox{$B\rightarrow B_*W$}
and \hbox{$W\rightarrow BW_*W$} in easy notations, where the star indicates the
place of neighbour~4. We assume here that the functions $f(\nu)$ 
and~$\sigma(\nu)$ are known. Their computation is efficient but it requires
notions which we have not mentioned here. We refer the reader 
to~\cite{mmJUCSii,mmbook1,mmbook2} for additional information. 

Clearly,
the coordinate of a neighbour~$N$ of a tile~$T$ with coordinate~$\nu$ depends on
the side~$\tau$ shared by~$T$ and~$N$. Now, notice that this side numbered by~$\tau$
in~$T$ does not receive the same number in~$N$ and we shall say that $N$~is
the neighbour~$\tau$ of~$T$. The correspondence between these
numbers gives the value of the function $t(\nu,\tau)$ and, for completeness,
we give it in Table~\ref{coord2}. Note that the sides of the central cell
are all numbered by~1{} in its neighbours. For the other cells, the correspondence
depends on the status of~$T$ and it may also depend on that of~$N$.
Side~7 is always the side shared by a neighbour which is on the same level of the
tree, even when there is a change of tree by the change of sector. If~$T$ is
black, its side~7 is numbered~2 on the other side. If~$T$ is white, the number
of its side~7{} in the other neighbour~$N$ depends on the status of~$N$ as indicated
in the table.

\def\lignetabii #1 #2 #3 #4 {%
\ligne{\hfill\hbox to 300pt{
\hbox to 50pt{\hfill#1\hfill}
\hbox to 50pt{\hfill#2\hfill}
\hbox to 50pt{\hfill#3\hfill}
\hbox to 50pt{\hfill#4\hfill}
}
\hfill}
}
\vtop{
\begin{tab}\label{coord2}
\leurre 
Correspondence between the numbers of a side shared by two heptagons,
$T$ and~$N$. Note that if $T$~is white, the other number of side~$1$
may be~$4$ or~$5$ when $N$~is white and that it is always~$5$ when
$N$~is black.
\end{tab}
\vspace{-8pt}
\grostrait
\ligne{\hfill\hbox to 100pt{\hfill black $T$\hfill}
\hfill
\hbox to 100pt{\hfill white $T$\hfill}\hfill}
\lignetabii {in $T$} {in $N$} {in $T$} {in $N$}
\vspace{-3pt}
\demitrait
\lignetabii 1 {3$^{wN}, $4$^{bN}$} 1 {4$^{wN}$, 5} 
\lignetabii 2      6     2     7 
\lignetabii 3      7     3     1
\lignetabii 4      1     4     1 
\lignetabii 5      1     5     1 
\lignetabii 6      2     6     2 
\lignetabii 7      2     7 {2$^{wN}$, 3$^{bN}$} 
\demitrait
\vskip 7pt
}

\def\lignetabiii #1 #2 #3 #4 #5 #6 #7 #8 {%
\ligne{\hfill\hbox to 340pt{
\hbox to 40pt{\hfill#1\hfill}
\hbox to 30pt{\hfill#2\hfill}
\hbox to 40pt{\hfill#3\hfill}
\hbox to 30pt{\hfill#4\hfill}
\hbox to 40pt{\hfill#5\hfill}
\hbox to 40pt{\hfill#6\hfill}
\hbox to 40pt{\hfill#7\hfill}
\hbox to 40pt{\hfill#8\hfill}
}
\hfill}
}
\vtop{
\begin{tab}\label{coord3}
\leurre 
The values of $v(\tau,\nu)$.
\end{tab}
\vspace{-8pt}
\grostrait
\lignetabiii {$\tau$} 1 2 3 4 5 6 7
\vspace{-3pt}
\demitrait
\lignetabiii {black} {$f(\nu)$} {$f(\nu)$$-$1} {$\nu$$-$1} {$\sigma(\nu)$} 
{$\sigma(\nu)$+1} {$\sigma(\nu)$+2} {$\nu$+1} 
\lignetabiii {left} {$f(\nu)$} {$\nu$$-$1} 
                    {$\sigma(\nu)$$-$$1$} {$\sigma(\nu)$} 
{$\sigma(\nu)$+1} {$\sigma(\nu)$+2} {$\nu$+1} 
\lignetabiii {white} {$f(\nu)$} {$\nu$$-$1} {$\sigma(\nu)$$-$1} {$\sigma(\nu)$} 
{$\sigma(\nu)$+1} {$\sigma(\nu)$+2} {$\nu$+1} 
\lignetabiii {right} {$f(\nu)$} {$\nu$$-$1} {$\sigma(\nu)$$-$1} {$\sigma(\nu)$} 
{$\sigma(\nu)$+1} {$\nu$+1} {$f(\nu)$+$1$} 
\lignetabiii {root} 0 1 {$\sigma(\nu)$$-$1} {$\sigma(\nu)$} {$\sigma(\nu)$+1} 
{$\nu$+1} 1
\vspace{-3pt}
\vspace{-3pt}
\demitrait
\vskip 3pt
\noindent
{\small\sc Note}: \leurre {\rm left} denotes the leftmost branch of the tree,
{\rm right} denotes its rightmost one. 
\vskip 2pt
In the case of {\rm left}, the sector for neighbours~$2$ and~$3$ is $\sigma\ominus1$.
In the case of {\rm right}, the sector for neighbours~$6$ and~$7$ is $\sigma\oplus1$.
\vskip 7pt
}
\vskip 5pt
   From this table, we can indicate the values of $v(\tau,\nu)$ which are given
in Table~\ref{coord3}. The basic point is that $v(1,\nu)$ for the heptagon~$H$
defined by~$\nu$ is always $f(\nu)$, as its neighbour~1 is the father of~$H$. 
Similarly, we have that $v(4,\nu)$ is always $\sigma(\nu)$ by definition,
regardless of the status of~$H$. Note that in the case 
of a black heptagon~$H$ on the leftmost branch of the tree, two of its neighbours 
belong to the other tree on this side of the sector of~$H$: neighbours~2 and~3.
Neighbour~2 is still \hbox{$\nu$$-$1} and, consequently, neighbour~3 
is the rightmost son of neighbour~2, hence it is 
\hbox{$\sigma(\nu$$-$$1)$+$1=\sigma(\nu$$-$$1)$}. 
A symmetrical remark holds for a white~$H$ standing on the rightmost branch of 
the tree: neighbours~6 and~7 belong to another tree, the one which spans the 
other sector than that of~$H$. Now, this time, neighbour~6 is 
numbered \hbox{$\nu$+1} and so, as neighbour~7 is the father of neighbour~6,
neighbour~7 is numbered \hbox{$f(\nu$+$1)=f(\nu)$+1}. At last, the root, which is a 
white node, belongs to both the left- and the rightmost branches of the tree. 
This is why it has a specific profile, different from both a standard white 
node and from a node on the rightmost branch of the tree. 

   It remains to indicate that in the case of a heptagon~$H$ which is on the
left- or the rightmost branch, it is easy to define the number of the sector
to which belongs the neighbours which do not belong to the tree of~$H$.
The new number is $\sigma\ominus1$ for neighbours~2 and~3 of a tile which is 
on the leftmost branch of the tree, it is $\sigma\oplus1$ for neighbours~6 and~7 
of a tile which is on the rightmost branch, see Table~\ref{coord3}.

\section{The triangular grids}
\label{trigrids}

   Now, we can define the new family of tilings constructed upon the tilings $\{p,q\}$.
This is the goal of Subsection~\ref{deftrigrids}. Then, in Subsection~\ref{coordtrigrids},
we construct coordinates for the tiles of the tilings which we define 
in Subsection~\ref{deftrigrids}.

\subsection{Defining the trigrids}
\label{deftrigrids}

   In~\cite{mmbook2}, the author gives a system of coordinates for the points of the
hyperbolic plane based on the heptagrid. The idea, illustrated by Figure~\ref{eclate_73},
is to iterate a simple construction.
In the first step, we split each heptagon~$H$ in seven equal triangles, each one based on
an edge of~$H$ and having its summit at the centre of~$H$. This first step of the 
process defines {\bf 1-triangles}. Now, considering that 
the 
{\bf $n$-triangles} have been defined, for each $n$-triangle~$T$, we obtain four
$n$+1-triangles as follows: three of these triangles are defined by a vertex~$V$ of~$T$
and by the mid-points of the two edges of~$T$ which meet at~$V$; the fourth $n$+1-triangle
is defined by the mid-points of the edges of~$T$. In Figure~\ref{eclate_73}, we can see
the 1-triangles and the 2-triangles obtained from a heptagon.

   We can describe this process more precisely and generalize it as follows.

   We fix a tiling~$\{p,q\}$ of the hyperbolic plane. Accordingly, each tile has a 
coordinate $(\sigma,\nu)$, with $\sigma$ the number of the sector where the tile
lies and $\nu$ its coordinate in the sector, the coordinate being~0 for the central 
tile.

   Each polygon~$P$ defines $p$~1-triangles, each one being numbered from~1 up to~$p$.
The 1-triangle~$i$ has the side~$i$ of~$P$ as its basis and the centre of~$P$ as the
vertex, opposite to the basis. In each triangle, we number the vertices from~0 to~2
with 2~being given to the centre of~$P$ and~0 with~1 associated to the ends of the
side~$i$. We fix the positions of~0 and~1{} in such a way that 0,1 and~2 are obtained
by counter-clockwise turning around the 1-triangle. Once the numbering of the vertices
are fixed, we use the same numbers to number the edges and the mid-points of the edges.
By convention, the edge~$i$ is opposite to the vertex~$i$ and the mid-point of the 
edge~$i$ is also called~$i$. Once the $n$-triangles are defined with the numbering
of their vertices, their edges and the mid-points of their edges are fixed, we can 
define the $n$+1-triangles as follows. Consider an $n$-triangle~$T$. If $\alpha$ is 
a vertex of~$T$, then the edges meeting at~$\alpha$ are numbered $\beta$ and~$\gamma$
with \hbox{$\{\alpha,\beta,\gamma\}=\{1,2,3\}$}. It is the same for the mid-points
of these edges and so, the vertex~$\alpha$ and the mid-points $\beta$ and~$\gamma$
define an $n$+1-triangle. Again by definition, the mid-points of the edges of~$T$
also define an $n$+1-triangle. Now, as the vertices of each $n$+1-triangle has received
\ligne{\hfill}
\vtop{
\ligne{\hfill
\scalebox{0.90}{\includegraphics{new_eclate_7_3.ps}}
\scalebox{0.36}{\includegraphics{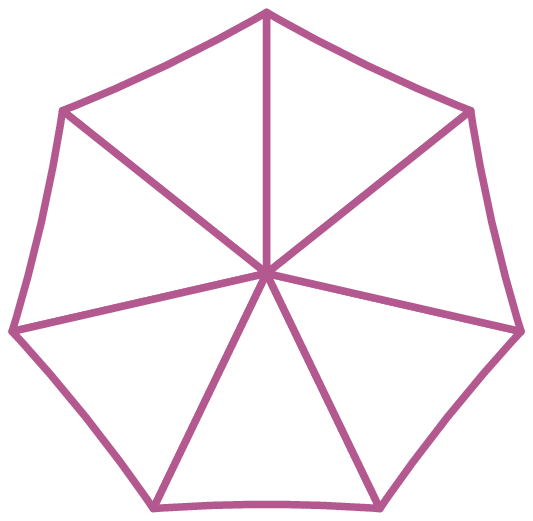}}
\hskip-60pt
\raise-15pt\hbox{\scalebox{0.315}{\includegraphics{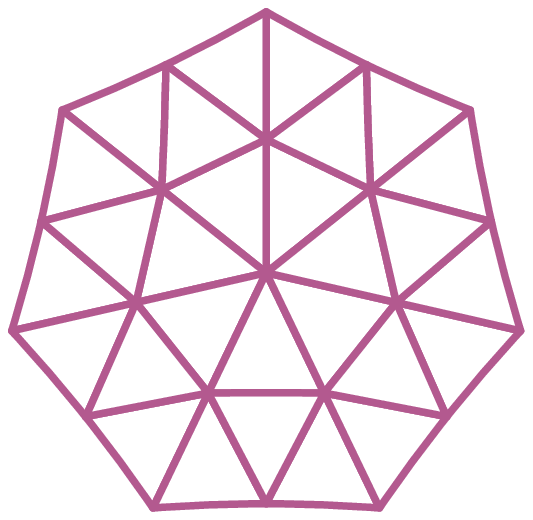}}}
\hfill}
\vspace{-10pt}
\begin{fig}\label{eclate_73}
From the heptagrid to the second triangular heptagrid, the heptatrigrid. 
\end{fig}
}

\noindent
a number in $[0..2]$ in such a way that we have 0,1, 2 or 0,2,1 while counter-clockwise
turning around the $n$+1-triangle, the edges and the mid-points of each $n$+1-triangle
can also be numbered by applying the same rules as those applied for numbering the
respective elements of the $n$-triangles. At last, let $T$ be an $n$+1-triangle~$T$
and let~$R$ be the single $n$-triangle which contains~$T$. If $T$~has as a vertex 
the vertex~$\alpha$ of~$R$, $T$~is numbered~$\alpha$. If all the vertices of~$T$
are the mid-points of the edges of~$R$, then $T$ is numbered with~3.

   Accordingly, this process can be repeated endlessly as it was completed for the
1-triangles. We can notice that if the numbering of an $n$-triangle~$T$ is \hbox{0, 1, 2} 
while counter-clockwise turning around~$T$, the numbering of the $n$+1-triangles
inside~$T$ numbered 0, 1 and~2 has the opposite orientation, while the numbering of
the $n$+1-triangle~3 has the same orientation.

   For each $n$, the set of $n$-triangles constitutes a triangular tiling of the hyperbolic
plane which we call the {\bf $n$-trigrid} based on the tiling $\{p,q\}$. As already 
mentioned in the captions, the trigrids based on the heptagrid are called 
{\bf heptatrigrids}.

\vskip 10pt
\vtop{
\ligne{\hfill
\scalebox{0.30}{\includegraphics{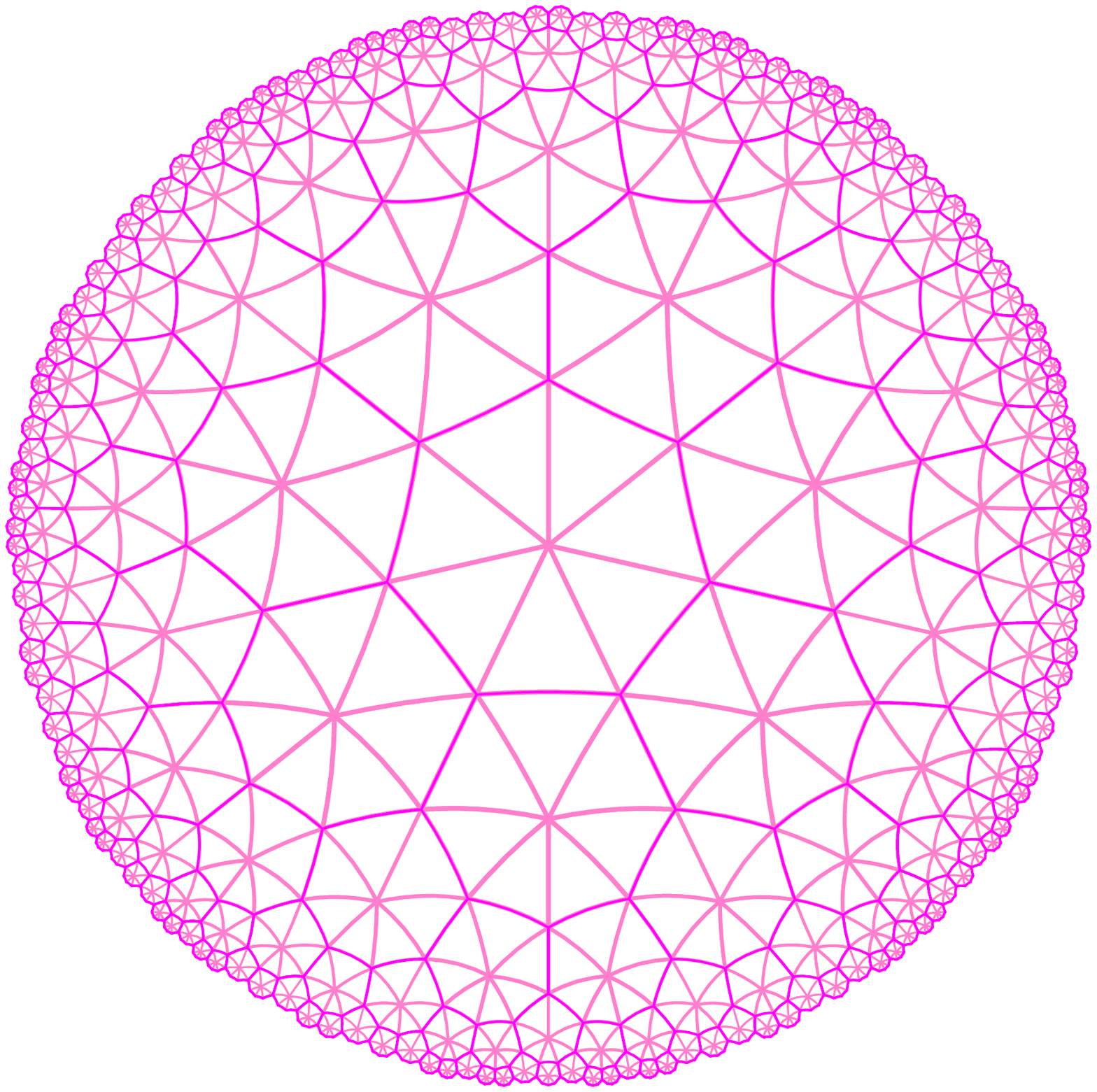}}
\scalebox{0.30}{\includegraphics{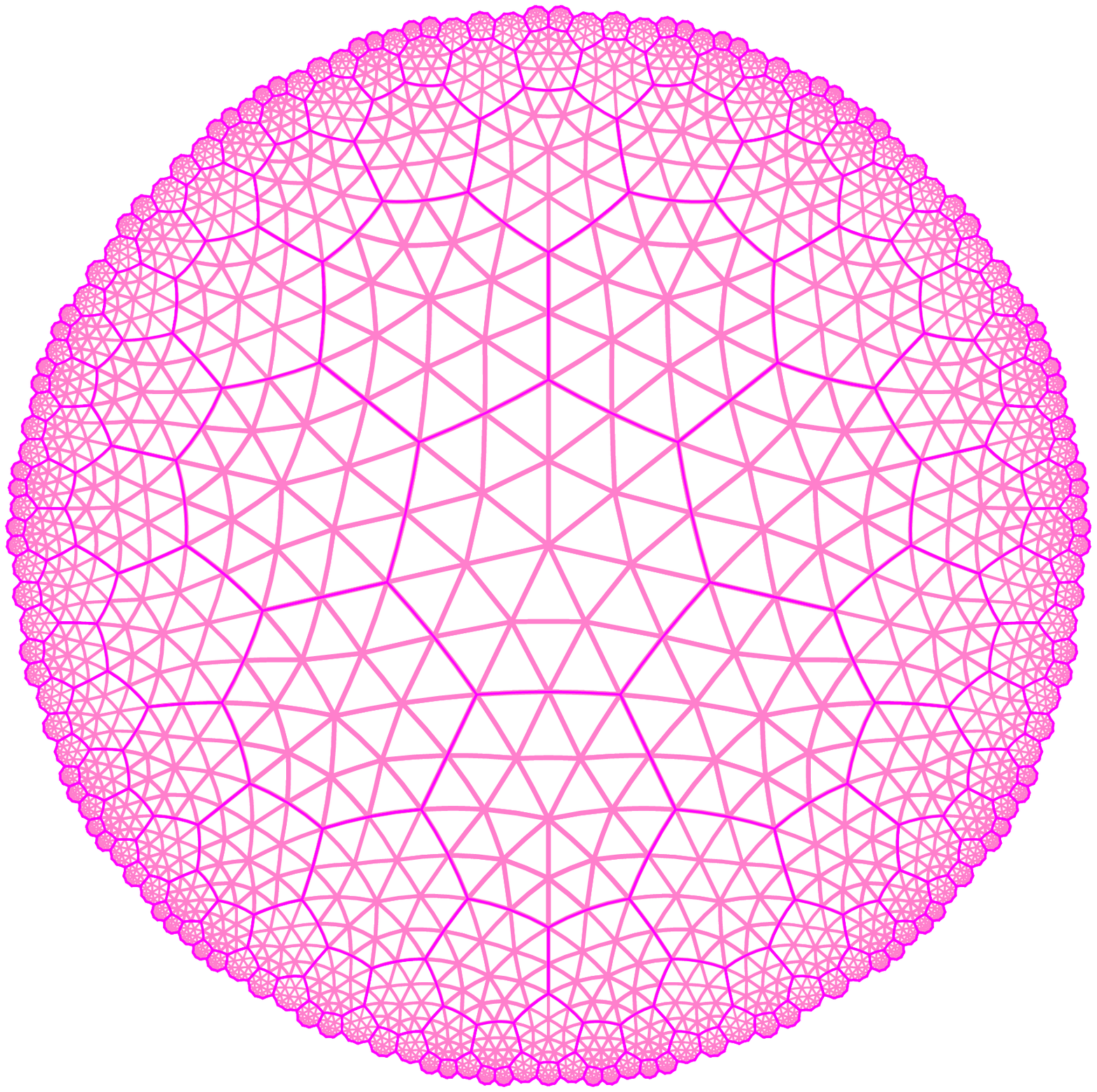}}
\hfill}
\begin{fig}\label{subtilings}
Two heptatrigrids: generation~$1$ on the left-hand side, generation~$2$
on the right-hand side.
\end{fig}
}

\subsection{Coordinates for the trigrids}
\label{coordtrigrids}

   Now, we have all the elements which allow us to define coordinates for the tiles of the
trigrids.

   Consider an $n$+1-triangle~$T$, with $n>0$. There is a unique sequence 
of~$n$+1 $m$-triangles $T_i$ with $m\leq n$, $T_{i+1}\subset T_i$ for $i \in [1..n]$,
with $T_{n+1}=T$ and each $T_i$ being an $i$-triangle. Let $[i]$~be the number of~$T_i$.
Then the coordinate of~$T$ is the sequence 
$(\sigma,\nu,[\alpha_1],..,[\alpha_{n+1}])$. We remark
that $[\alpha_1]\in[1..p]$ and that $[\alpha_i]\in[0..3]$ for $i\in[2..n$+$1]$.

   In~\cite{mmbook2} the author defined coordinates of the points of the hyperbolic plane
starting from the heptagrid by considering, for each point, a sequence of $n$-triangles
converging to the point. The numbers of the $n$-triangles which are defined 
in~\cite{mmbook2} exactly in the same way as they are defined in 
Subsection~\ref{deftrigrids} are used as digits for the representation of the point. 
There are two big differences with the situation we consider in this paper: 
in~\cite{mmbook2}, the coordinate of a point is an infinite sequence while here, it is
a finite one; the second difference is that a point has infinitely many possible
coordinates while here, an $n$-triangle has a unique coordinate.

    In this sub-section, we shall see how to compute the coordinates of the neighbours
of a tile.

    Consider an $n$+1-triangle~$T$. Let $S$~be the $n$-triangle which contains~$T$.
We shall see that if we know the coordinates of the neighbours of~$S$, we can easily
compute the coordinates of the neighbours of~$T$.

    The reason will be clear from the following study.

    If~[3] is the number of~$T$, all its neighbours are in~$S$. in the other cases,
one neighbour is in~$S$ while the others are not, but, as can easily be seen, they
are necessarily in a neighbour of~$S$. We denote the neighbours of an $n$-triangle
in a way which similar to what we did for a polygon. Here, the neighbours~1, 2 and~3
of an $n$-triangle~$Z$ are the $n$-triangles which share with~$Z$ the side~0, 1 and~2
respectively. The exact correspondence for the neighbours of~$T$ is given by the
following table in which $U$, $V$ and~$W$ are the neighbours~1, 2 and~3 respectively
of~$S$. Also, we denote by $<$$Z$$>$ the coordinate of the $n$-triangle~$Z$ and
the $n$+1-triangles contained in~$Z$ have $<$$Z$$>$.[0], $<$$Z$$>$.[1],
$<$$Z$$>$.[2] and $<$$Z$$>$.[3] as coordinates, where '.' denotes the concatenation
which transforms a sequence with~$k$ terms into a sequence with~$k$+1 terms.

\def\neighbourrow #1 #2 #3 #4 {%
\ligne{\hfill
\hbox to 70pt{\hfill#1\hfill}
\hbox to 50pt{\hfill#2\hfill}
\hbox to 50pt{\hfill#3\hfill}
\hbox to 50pt{\hfill#4\hfill}
\hfill}
}

\vtop{
\begin{tab}\label{coordneigh}
\leurre
Table of the coordinates of the neighbours of an $n$$+$$1$-triangle~$T$,
knowing the coordinates of the neighbours of the $n$-triangle~$Z$ which
contains~$T$ and also knowing the coordinates of the neighbours of~$Z$.
\end{tab}
\vspace{-12pt}
\ligne{\hfill
\vtop{\leftskip 0pt\parindent 0pt\hsize=250pt
\grostrait
\neighbourrow {$Z$} {$U$} {$V$} {$W$}
\vspace{-6pt}
\demitrait
\vspace{4pt}
\neighbourrow {$T$.0} {$<$$T$$>$.[3]} {$<$$W$$>$.[1]} {$<$$V$$>$.[1]}
\neighbourrow {$T$.1} {$<$$W$$>$.[0]} {$<$$T$$>$.[3]} {$<$$U$$>$.[0]}
\neighbourrow {$T$.2} {$<$$V$$>$.[2]} {$<$$U$$>$.[2]} {$<$$T$$>$.[3]}
\neighbourrow {$T$.3} {$<$$T$$>$.[0]} {$<$$T$$>$.[1]} {$<$$T$$>$.[2]}
\vspace{-4pt}
\demitrait
\vskip 7pt
}
\hfill}
}

   From this table, we can easily devise an algorithm which computes the coordinates
of the neighbours of a tile~$T$. The idea is simple: we start with computing the 
neighbours $U_1$, $V_1$ and~$W_1$ of the 1-triangle~$T_1$ which contains~$T$. 
We know that $U_1$ and~$V_1$ belong to the same polygon~$P$ as~$T_1$, they are 
simply based on another side of~$P_1$. More precisely, if~$\tau$ is the number of the side
on which $T_1$~is based, then $U_1$~is based on the side~$\tau\oplus1$ and $V_1$~is based
on the side~$\tau\ominus1$. Now, $W_1$ belongs to a polygon~$Q$ which is the 
neighbour~$\tau$ of~$P$ and the number in~$Q$ of the side which it shares with~$P$ is
given by a table similar to Table~\ref{coord2} which can be used in the case of a
heptatrigrid. Next, we go from the neighbours $U_m$, $V_m$, $W_m$ of~$T_m$, where
$T_m$ is the $m$-triangle which contains~$T$, to those, $U_{m+1}$, $V_{m+1}$,
$W_{m+1}$ of~$T_{m+1}$ thanks to Table~\ref{coordneigh}. The exact writing of this
algorithm in a pseudo-code or in a programming language is left to the reader.

   Now, we can notice that each step of the considered algorithm is constant in term
of resources and in term of what is appended to the already known resources. Accordingly,
this algorithm is linear. Now, it was proved in~\cite{mmbook2} that the computation
of the coordinates of the neighbours of a tile in a tiling $\{p,q\}$ is linear in the
size of the coordinate of the tile itself. Accordingly, this proves the following
result:

\begin{thm}\label{coordlin}
For each $p$ and $q$ such that the tiling $\{p,q\}$ is a tiling of the hyperbolic plane,
there is an algorithm which computes the coordinates of the neighbours of an 
$n$-triangle~$T$ in linear time with respect to the size of the coordinate of~$T$.
\end{thm}

   It is worth noticing that this algorithm allows us to compute the size
of the coordinates of a tile for each $n$-trigrid, independently of~$n$. Note that,
in any case, the size of~$n$ is contained in unary in the size of the coordinate
as the number of numbers of $m$-triangles involved in the coordinates is $n$$-$1.
But, of course, there is an algorithm for each $p$ and~$q$.

\section{Conclusion}

    With Theorem~\ref{coordlin}, we have an algorithm which allows us to efficiently
compute the coordinates of the neighbours of a tile in any $n$-trigrid of a given
tiling~$\{p,q\}$. This allows us to implement cellular automata in such tilings.
In particular, this holds in the heptatrigrid.

    It might be objected that the resulting cellular automaton has not a uniform
structure. This is not completely true. First of all, each triangle has three neighbours
so that, as long as the neighbours are defined by the condition of sharing a side with
the tile, the uniformity on the number of neighbours is preserved. What is here 
different is that the number of tiles around a vertex is not always the same. It 
is~6{} in the majority of cases but it is~$p$ in the others: in fact, when an 
$m$-triangle has the centre~$C$ of a polygon as a vertex, there are $p$ $m$-triangles 
around~$C$, regardless of~$m$. But, this occurrence of vertices around which there is 
a number of sides which is different of~6 is also regular as the tiling $\{p,q\}$ itself is
regular, being defined by the tessellation of a regular polygon.

    Such an implementation of cellular automata was done in the heptatrigrid,
see~\cite{mmarXivbacteria}.

\end{document}